\documentclass[%
 reprint,
superscriptaddress,
nofootinbib,
 amsmath,amssymb,
prl
]{revtex4-2}
\usepackage{graphicx} %
\usepackage{amsmath,amssymb}
\usepackage{todonotes}
\usepackage{hyperref}
\usepackage{svg}

\DeclareMathOperator{\SU}{SU}
\DeclareMathOperator{\U}{U}
\DeclareMathOperator{\SO}{SO}
\DeclareMathOperator{\Spin}{Spin}
\DeclareMathOperator{\OO}{O}
\DeclareMathOperator{\Tr}{Tr}

\date{August 2024}

\begin{document}
\title{Emergent Non-Invertible Symmetries Bridging UV and IR Phases\\ --- The Adjoint QCD Example ---}

\author{Michele Del Zotto}
\affiliation{Mathematics Institute, Uppsala University, Box 480, SE-75106 Uppsala, Sweden}
\affiliation{Center For Geometry and Physics, Uppsala University, Box 480, SE-75106 Uppsala, Sweden}
\affiliation{Department of Physics and Astronomy, Uppsala University, Box 516, SE-75120 Uppsala, Sweden}
\author{Shani Nadir Meynet}
\author{Daniele Migliorati}
\affiliation{Mathematics Institute, Uppsala University, Box 480, SE-75106 Uppsala, Sweden}
\affiliation{Center For Geometry and Physics, Uppsala University, Box 480, SE-75106 Uppsala, Sweden}
\author{Kantaro Ohmori}
\affiliation{Department of Physics, University of Tokyo, Hongo 7-3-1, Bunkyo, Tokyo, Japan}

\begin{abstract} 

    In this letter, we demonstrate how an emergent non-invertible symmetry along a renormalization group (RG) flow reveals connections between microscopic and macroscopic physics.
    We illustrate this using (3+1)-dimensional Adjoint QCD with two flavors of Weyl fermions as an example.
    For the $\SU(2)$ case, C\'ordova and Dumitrescu proposed a non-supersymmetric deformation of the $\mathcal{N}=2$ SYM theory leading to dynamical abelianization, followed by monopole condensation, and resulting in a confining infrared (IR) phase characterized by disjoint copies of the $\mathbb{CP}^1$ sigma model.
    In this scenario, we point out that the abelianized phase has an emergent non-invertible symmetry, which is matched with the non-invertible symmetry of the IR $\mathbb{CP}^1$ phase, associated to the Hopf solitons.  This result illustrate how an emergent non-invertible symmetry can be used to provide a bridge connecting the IR solitons and their properties with the ones of microscopic degrees of freedom in gauge theories with one-form symmetries. Moreover, based on this insight we generalize these results to other gauge theories with any number of colors, and propose a candidate for the UV baryon operator in all these cases.

\end{abstract}

\maketitle

In four-dimensional gauge theories Wilson and 't Hooft loops 
are well-known order parameters for confinement \cite{Wilson:1974sk,tHooft:1977nqb}%
, related to the spontaneous breaking of one-form symmetries %
\cite{Gaiotto:2014kfa}, %
which 
significantly advanced our understanding of the dynamics of these %
models (see e.g.\ \cite{Aharony:2013hda,Kapustin:2014gua,Gaiotto:2017yup}). %
Recently, generalized symmetries of a different type---non-invertible symmetries---have been studied in theories in various dimensions  (see e.g.~\cite{Cordova:2022ruw,McGreevy:2022oyu,Gomes:2023ahz,Schafer-Nameki:2023jdn,Brennan:2023mmt,Bhardwaj:2023kri,Shao:2023gho,Carqueville:2023jhb,Iqbal:2024pee} for %
reviews). 
As non-invertible %
symmetries in 3+1 dimensions %
arise from 
mixtures of 0-form and 1-form symmetries, it is natural to explore %
their implications 
to 
confinement. 
One intriguing scenario is \textit{dynamical abelianization} \cite{tHooft:1981bkw}, where a non-abelian gauge 
theory is approximated by an abelian gauge theory above %
the confining scale. 
The abelian phase can potentially have %
infinite non-invertible %
symmetries 
\cite{Cordova:2022ieu,Choi:2022jqy} with non-trivial higher %
structures \cite{Copetti:2023mcq}, 
suggesting a possible match with further infrared (IR) phases where confinement occurs.

In this note, we propose an example of this scenario in 
four-dimensional adjoint QCD with two 
Weyl fermion flavors ($\text{AdQCD}_{4}^{(2)}$). As highlighted by C\'ordova and 
Dumitrescu \cite{Cordova:2018acb}, %
a deformation of $\mathcal{N}=2$ \(\SU(2)\) Super-Yang-
Mills (SYM) theory enables us to examine the scenario of dynamical abelianization.%
\footnote{$\text{AdQCD}_{4}^{(2)}$ and its abelianization can also be studied via circle compactification, see e.g. \cite{Unsal:2007jx}. Also the relation between such method and generalized symmetry is discussed, e.g.\ in \cite{Anber:2018iof}.}

Among the possible IR phases 
of $\text{AdQCD}_{4}^{(2)}$, there is a confining one, %
expected 
to be described by disjoint copies of the $
\mathbb{CP}^1$ sigma model. The latter exhibits a non-invertible symmetry, %
described by Chen--Tanizaki 
and Hsin 
\cite{Chen:2022cyw,Hsin:2022heo,Chen:2023czk,Pace:2023kyi,Pace:2023mdo}.%
In this letter we 
demonstrate a matching of 
the non-invertible symmetry between the abelianized 
phase and the confined phase. %
In the $\SU(2)$ case, this matching of non-invertible symmetry is the precise description of what C\'ordova and Dumitrescu proposed in \cite{Cordova:2018acb}.
Additionally, 
we draw physical insights from this refined understanding and 
we generalize the 
discussion to other gauge theories with a general 
number of colors. %

Our study demonstrates that non-invertible symmetries provide a bridge connecting %
IR solitons 
and their properties 
with the ones of microscopic degrees of freedom, offering a %
cleaner 
picture helping to clarify
several longstanding open conjectures about the 
dynamics of these models. 
We illustrate the power of this technique in the $\text{AdQCD}_{4}^{(2)}$ case, but the logic we present is more general and we expect many similar applications for other QFTs and other dimensions.

\medskip

\noindent \textbf{Adjoint QCD and $\mathcal{N}=2$ Super Yang-Mills.} A convenient approach to study $\text{AdQCD}_{4}^{(2)}$  
consists of considering a deformation of %
$\mathcal{N}=2$ pure 
SYM%
--- 
a soft mass $M_\text{soft}$ %
for 
the $\mathcal N=2$ vector multiplet 
scalar $\phi$ breaking the whole of supersymmetry (see eg. \cite{Alvarez-Gaume:1996vlf,Luty:1999qc,Cordova:2018acb})
---and capitalize on 
the Seiberg-Witten (SW) solution \cite{Seiberg:1994rs,Seiberg:1994aj} %
.
The SW solution provides an effective field theory (EFT) 
description on the Coulomb branch of the supersymmetric theory 
where %
$\phi$ takes a generic vev, abelianizing the theory.

Let us begin by considering the $\SU(2)$ case. Crucial to the SW solution in this
case are two points along the Coulomb branch where a monopole and a dyon 
respectively become massless. In a suitable electromagnetic duality frame, 
each such point is  described by a $\U(1)$ $\mathcal N=2$ quantum electrodynamics (SQED) 
involving a single charge one hypermultiplet.

C\'ordova and Dumitrescu proposed that, with the $\mathcal{N}=0$ deformation, 
the hypermultiplet scalar $h_i$ at the monopole or dyon point %
takes a vev %
breaking $\SU(2)_R$ to $\U(1)$ 
($i=1,2$ is an $\SU(2)_R$ index)
. Such a vev in the abelian gauge theory 
leads to an IR nonlinear sigma model with target $\mathbb{CP}^1 = \SU(2)/\U(1)$.
Actually, each of the monopole and dyon point must lead to a disjoint copy of $\mathbb{CP}^1$, 
thus there are two vacua breaking $(\mathbb{Z}_8)_R$, ie. the non-anomalous part of the $\U(1)_R$ 
symmetry of $\mathcal N=2$ SYM, and each vacua is described by a $\mathbb{CP}^1$ model in the IR. 
We denote this IR theory $(\mathbb{CP}^1)^{\sqcup 2}$.

If we take the limit of $M_\text{soft}\to \infty$, the UV theory becomes precisely $\text{AdQCD}_{4}^{(2)}$, 
ie. $\mathcal{N}=0$ $\SU(2)$ QCD with $n_f=2$ adjoint fermions $\lambda_i$ ($i=1,2$).
The R-symmetry $\SU(2)\times \mathbb{Z}_8$ of $\mathcal{N}=2$ SYM, acting on $\lambda_i$, 
is preserved by $M_\text{soft}$. If we further assume that the fermion bilinear condenses:
\begin{equation}
    \langle \Tr(\lambda_i \lambda_j) \rangle \neq 0,
\end{equation}
the symmetry $(\SU(2)\times \mathbb{Z}_8)/\mathbb{Z}_2$ breaks down to  $\OO(2)$, 
predicting the IR phase $(\mathbb{CP}^1)^{\sqcup 2}$, which is exactly 
the endpoint of the RG flow of deformed $\mathcal{N}=2$ SYM.

\begin{figure}[t]
    \centering
    \includeinkscape[width=.9\linewidth]{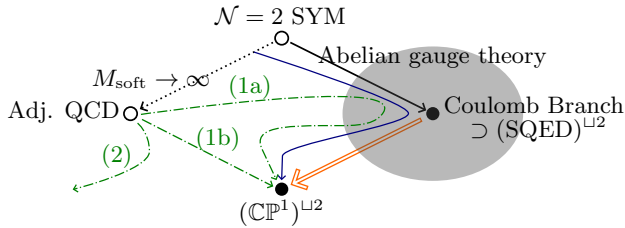}
    \caption{Cartoon of RG flows around $\mathcal{N}=2$ SYM and $\text{AdQCD}_{4}^{(2)}$. See the main text for explanation.
    } 
    \label{fig:RG_cartoon}
\end{figure}
The RG flows relevant in this context are summarized in Fig.~\ref{fig:RG_cartoon}. 
The black solid straight line is the SW flow to the Coulomb branch theory, 
which %
contains the disjoint copies of SQEDs for %
the monopole and dyon points. 
The grey region is where the theory admits an abelian gauge theory description. 
The blue solid curve is the C\'ordova-Dumitrescu flow. 
The dotted straight line is the $M_\text{soft}$ deformation from SYM to $\text{AdQCD}_{4}^{(2)}$.
For $\text{AdQCD}_{4}^{(2)}$, we have three potential scenarios, %
the dashed curves in Fig.~\ref{fig:RG_cartoon}.  
The first scenario, (1a), assumes that $\text{AdQCD}_{4}^{(2)}$ is abelianized, 
before the chiral symmetry breaking. 
The abelianization occurs by condensing the (gauge variant) 
fermion bilinear $\sum_i \lambda_i^2$. Note that, while in SYM the abelianization 
is caused by condensing $\phi$, the $M_\text{soft}$ deformation 
can mix the scalar $\phi$ and the bilinear $\sum_i \lambda_i^2$, 
and thus it is possible that the two regimes are continuously connected. 
The scenario (1b) is continuously connected to (1a), but in such a scenario 
there is no clear abelianization along the flow to $(\mathbb{CP}^1)^{\sqcup 2}$. 
Even in this scenario, (generalized) symmetry based analysis for 
the flow studied by C\'ordova-Dumitrescu provides a qualitative 
matching between the IR and UV physics. 
The scenario (2) is disconnected from the others and corresponds to 
flows of $\text{AdQCD}_{4}^{(2)}$ to phases that differ from $(\mathbb{CP}^1)^{\sqcup 2}$. 
In this case we expect a phase transition before $M_\text{soft}=\infty$. 
We cannot analytically know which scenario might be realized. 
We %
explain the flow corresponding to the orange hollow arrow below.

The preceeding discussion generalizes to the $N_c > 2$  cases \cite{DHoker:unpublished,Nardoni:talk,DumitrescuNardoni:talk} -- building upon \cite{DHoker:2020qlp,DHoker:2022loi}.  In this letter we provide a brief overview, along with a perspective informed by non-invertible symmetry. The $\mathcal N=2$ $\SU(N_c)$ SYM theory has $N_c$ points where $N_c -1$ monopoles (or dyons), 
all mutually local, become simultaneously massless \cite{Argyres:1994xh,Klemm:1994qj,Douglas:1995nw}. %
These points are rotated by the non-anomalous $\mathbb Z_{4N_c}$ 
subgroup of the $\U(1)_R$ symmetry, acting on the Coulomb branch coordinates 
as a $\mathbb Z_{N_{c}}$. The theory at each of those points can be described by $N_c -1$ 
copies of massless SQED with a unit charge hyper (for the dual photons) \cite{Douglas:1995nw}.\footnote{This might suggest that a subgroup 
of the Weyl group of 
$SU(N_c)$ 
is restored at this point, leading to a more interesting topological order. In such case, an analysis 
as a semi-abelian theory 
would be more precise \cite{Nguyen:2021yld}.} On the other hand, a possible IR phase of %
$\text{AdQCD}_{4}^{(2)}$, 
which we denote $(\mathbb{CP}^1)^{\sqcup N_c}$, 
is given by $N_c$ copies of a $\mathbb{CP}^1$ model 
(conjecturally realized via condensation of fermion bilnears---see \cite{Shifman:2013yca} for a review). %
As discussed in more detail below, 
building on the matching of symmetries we expect that upon turning on 
$M_{\text{soft}}$ 
each multimonopole point contributes a $\mathbb{CP}^{1}$ factor to the IR phase, thus realizing a version of the scenarios (1a) and (1b) also 
in the $N_c >2$ cases. This is indeed consistent and confirmed by the D'Hoker, Dumitrescu, Nardoni and Gerchkovitz analysis \cite{DHoker:unpublished}.

\medskip

\noindent \textbf{Emergent non-invertible chiral symmetries in SQED.} The $\mathcal{N}=2$ Coulomb branch theory at the monopole point is described by a
collection of $N_c-1$ vectormultiplets $(\tilde{A}_k,
\rho_k,\varphi_k)$ giving a $\prod_{k=1}^{N_c-1} \U(1)_{k,\text{mag}}$ gauge 
symmetry and a collection of $N_c - 1$ hypermultiplets 
$(h_{i,k},\overline{h}_{i,k}, \psi_{\pm,k})$ each with unit 
charge with respect to a single $\U(1)_{k,\text{mag}}$ factor.
The scalars $\varphi_k$ are identified with the dual special coordinates $a_{D,k}$ 
on the Coulomb branch that vanish at the monopole point.
The matter content is summarized in Table~\ref{tab:SQED}. \begin{table}[t]
    \centering
    \begin{tabular}{c|c|c|c|c|c|c|c}
          & $\tilde{A}_k$ & $\rho_{i,k}$ & $\varphi_k$ & $h_{i,k}$ &$\overline{h}_{i,k}$ & $\psi_{k,+}$ & $\psi_{k,-}$ \\ \hline
         spin& 1 & $\frac12$ & 0 & 0 & 0 &$\frac12$ & $\frac12$  \\
         $\U(1)_{\ell,\text{mag}}$ & 0 & 0 & 0 & $\delta_{k,\ell}$ & $-\delta_{k,\ell}$ & $\delta_{k,\ell}$ &$ -\delta_{k,\ell}$ \\
         $\U(1)_X$ & 0 & -1 & 2 & 0 & 0 & 1 & 1 \\
         $\SU(2)_R$ & $\mathbf{1}$ & $\mathbf{2}$ & $\mathbf{1}$ & $\mathbf{2}$ & $\mathbf{2}$ & $\mathbf{1}$ & $\mathbf{1} $
    \end{tabular}
    \caption{Field content of the SQED at the multimonopole point.}
    \label{tab:SQED}
\end{table}
The symmetry $\U(1)_X$, which is the superconformal $\U(1)_R$ symmetry for the free 
limit of the theory, is emergent at this point.
In addition%
, the SQEDs have also one-form symmetries $\U(1)^{(1)}_k$, which are magnetic with respect to the EFT gauge field $\tilde{A}_k$, and electric to the original gluon. The UV theory of the deformed $\mathcal{N}=2$ SYM has a $\mathbb{Z}_{N_c}^{(1)}$ electric one-form symmetry, %
identified with a $\mathbb{Z}_{N_c}$ subgroup of $\U(1)^{(1)}_\text{diag} \subseteq \prod_k \U(1)^{(1)}_k$. The emergent $\U(1)_X$ symmetry has ABJ anomalies with each factor of the abelianized gauge group. In particular, for reasons that will become clear %
below, considering the diagonal gauge transformations $\U(1)_\text{diag}$ we obtain the following ABJ equation:
\begin{equation}
    \partial_\mu j^\mu_X = \frac{N_c-1}{4\pi^2}\epsilon_{\alpha\beta\gamma\delta} \tilde F^{\alpha\beta}_\text{diag}\tilde F^{\gamma\delta}_\text{diag}.
    \label{eq: anomaly}
\end{equation}
According to \cite{Choi:2022jqy,Cordova:2022ieu}, this induces non-invertible symmetries, corresponding to $\U(1)_X$ rotations (up to an invertible $\mathbb Z_{2(N_c-1)}$ %
subgroup). In particular, the topological operator $\mathcal{D}_\alpha$ corresponding to an $\alpha= 2\pi\frac{p}{q}$ rotation, placed on a 3-dimensional submanifold $M$ of the spacetime is constructed by stacking the ``naive" topological operator generating the 0-form symmetry $\U(1)_X$ with a TQFT:
\begin{equation}
    \mathcal{D}_\alpha = \exp\left(\mathrm{i}\alpha\int_M j_X\right) \mathcal{A}^{q,2(N_c-1) p}[\tilde{A}],
    \label{eq: Dcal}
\end{equation}
where $\mathcal{A}^{q,p}$ is the minimal abelian TQFT introduced by \cite{Hsin:2018vcg}, coupled to the bulk gauge field $\tilde{A}_\text{diag}$.\footnote{When $q$ and $2(N_c-1)p$ are coprime, we understand $\mathcal{A}^{q,2(N_c-1)p}$ as $\mathcal{A}^{q/g,2(N_c-1)p/g}$ with $g$ being the greatest common divisor.}
This operator has the following fusion rule with its adjoint, mixing 0-form and 1-form symmetries:
\begin{equation}
    \mathcal{D}_{\alpha}^* \mathcal{D}_\alpha [M] = \text{condensation operator on $M$},
\end{equation}
where the right hand side is the operator on $M$ obtained by condensing a $\mathbb{Z}_{q/\mathrm{gcd}(q,N_c)}$ subgroup of $\U(1)^{(1)}$. For a review about the  condensation operators, we refer our readers to  \cite{Roumpedakis:2022aik}.%
\footnote{\ There are also other constructions %
allowing 
topological defects corresponding to 
irrational chiral rotations \cite{GarciaEtxebarria:2022jky,Karasik:2022kkq,Arbalestrier:2024oqg,Hasan:2024aow}. Our results are not affected by such choices.}

\medskip

\noindent \textbf{Gauge theories and $\mathbb{CP}^1$ phases.} In the $N_c = 2$ case, 
the potential induced from the UV to the effective SQED by matching the stress-energy tensor multiplet
is \cite{Luty:1999qc,Cordova:2018acb}
\begin{equation}
V = V_0 + M^2_\text{soft} \left( \frac{1}{e^2} \lvert\varphi \rvert ^2 - \frac{1}{2} \overline{h}_i h_i + \frac{\Lambda \mathop{\mathrm{Im}}\varphi}{ \pi^2} \right),
\label{eq:potential}
\end{equation}
where $e$ is the photon coupling and $\Lambda$ is the dynamical scale, $V_0 = \frac{e^2}{2} \left( \overline{h}_i h_i \right)^2 + 2\lvert\varphi \rvert ^2 \overline{h}_i h_i$ is the potential in the undeformed supersymmetric theory, obtained from the SW prepotential \cite{Seiberg:1994rs,Seiberg:1994aj}. 

In the actual SW solution, the coupling $e$ at the monopole point, which depends on $\Lambda$, is not small. However, C\'ordova and Dumitrescu argued that analysing the semi-classical region of $e \ll 1$ is still valuable for understanding the qualitative behaviour, in particular matching symmetries among the scales. 
We follow this same idea and assume $e \ll 1$. When $e^2\Lambda \ll M_\text{soft} \ll \Lambda$, the monopole hypermultiplet scalars develop the vev $\langle h_i\overline{h}_i \rangle \neq 0$, flowing to the $\mathbb{CP}^1$ model.
Recalling that we also have the dyon point related by the $\mathbb{Z}_8$ symmetry, %
the IR phase of the deformed SYM is expected to be $(\mathbb{CP}^1)^{\sqcup 2}$, and this guess extends to $\text{AdQCD}_{4}^{(2)}$.
Dynamical questions aside, in fact this argument connects $\mathcal{N}=2$ SYM, $\text{AdQCD}_{4}^{(2)}$ and $(\mathbb{CP}^{1})^{\sqcup 2}$ by a continuous deformation, consisting of dynamical RG flows as well as explicit tuning deformations of parameters. %
All the symmetries preserved by the deformations (and their 't Hooft anomalies) must match %
, %
as 
checked %
by C\'ordova and Dumitrescu \cite{Cordova:2018acb}.

How to extend this argument for $N_c > 2$? Here the monopole and dyon points are replaced by 
the multimonopole points \cite{DHoker:2020qlp,DHoker:2022loi,Nardoni:talk,DHoker:unpublished}, and the matching of symmetries is more subtle. Naively, at each multimonopole 
point, the UV $\SU(2)_R$ symmetry of $\mathcal N=2$ SYM enances to $\SU(2)^{N_c -1}$. Accordingly, a naive generalization of the C\'ordova-Dumitrescu scenario to the $N_c>2$ case is to get a direct product of $N_c-1$ copies of $\mathbb{CP}^1$ from each mutimonopole point.\footnote{\ The readers should not confuse this theory, which is a tensor product, with the theory we denote $(\mathbb{CP}
^{1})^{\sqcup N_c}$, which is a direct sum of disconnected vacua.} However, the direct product theory has relevant deformations respecting the UV symmetry but breaking $\SU(2)^{N_c-1}$ down to its diagonal.  With a choice of signs such deformations lift all these vacua except for the diagonal of the direct product, resulting in the expected $\mathbb{CP}^1$ phase---see the End Matter for a more detailed discussion about this point. %
This also implies that all the 1-form 
symmetries in the SQED description decouple but $\U(1)^{(1)}_\text{diag}$ of equation \eqref{eq: anomaly}. In 
particular, $\U(1)^{(1)}_\text{diag}$ crucially contains the unbroken UV center 1-form 
symmetry as a subgroup and so can be exploited to build a dictionary between the UV phase and 
the IR $\mathbb{CP}^1$: vortices carry color flux \cite{Bolognesi:2006ws,Bolognesi:2007ut}. 
In summary, also for $N_c>2$ we expect a version of scenarios (1a) and (1b) such that the 
continuously connected theories are $\mathcal{N}=2$ SYM, $\text{AdQCD}_{4}^{(2)}$, and $(\mathbb{CP}
^{1})^{\sqcup N_c}$.

\medskip

\noindent \textbf{Matching non-invertible symmetries.} Consider again the $N_c =2$ case. The $\U(1)_X$ symmetry in SQED acts on fermions, which decouple due to the Yukawa couplings
\begin{equation}
    h_i \rho_i \psi_- + \overline{h}_i \rho_i \psi_+.
    \label{eq: yukawa}
\end{equation}
However, if the sigma model scalars $h_i, \overline{h}_i$ ($\sum_i h_i \overline{h}_i= v^2$) have a topologically nontrivial solitonic profile over spacetime, such decoupled fermions can provide a coupling between the symmetry and the soliton. 
For the model where $\U(1)_\text{mag}$ is not gauged the effective action induced by the decoupled fermion is \cite{Abanov:1999qz,Hsin:2020cgg}:
\begin{equation}
    - \mathrm{i} \int A_X \wedge \mathcal{N}(h)\,,  \label{eq: effective charge}
\end{equation}
where $A_X$ is the $\U(1)_X$ background and $\mathcal{N}(h)$ is the soliton number density. The sign in front is our convention about the definition of signed number of solitons. 
The Skyrmion, the topological soliton corresponding to the wrapping number over $S^3$, \textit{carries $\U(1)_X$ charge induced by the decoupled fermions}. Another perspective on this effect is to consider the local operator creating a Skyrmion in the IR sigma model. Such an operator corresponds to an instanton-like profile of $h$ in which $h$ vanishes at a point in spacetime. With such a profile, the fermions with Yukawa coupling \eqref{eq: yukawa} have a zero mode around such an operator \cite{Cordova:2019jnf}---see \cite{Choi:2022odr} for the relationship to the viewpoint of \eqref{eq: effective charge}---and thus we have to insert fermions to saturate the corresponding zero mode and get a nonzero correlation function. Conversely, the one-point function of SQED operators $\rho_i$ and $h_i \psi_-$ is nonzero only if the Skyrmion number changes from the initial to final state, and thus we expect that such operators flow to operators creating/annihilating the soliton.

Upon gauging $\U(1)_\text{mag}$, we have a different soliton, the Hopf soliton, associated to the homotopy group $\pi_3(\mathbb{CP}^1)\cong \mathbb{Z}$. 
We can use \eqref{eq: effective charge} just by replacing $S^3$ by $\mathbb{CP}^1$ in this case, and thus we conclude that the Hopf soliton has a unit $\U(1)_X$ charge, negative in our convention. 
In particular, because of the spin-$\U(1)_X$ charge relation in SQED, the Hopf soliton is a fermion. 
This statistics of the soliton requires a discrete theta angle $\theta = \pi$ \cite{Witten:1983tx,Cordova:2018acb}.

Now, recall that $\U(1)_X$ is non-invertible due to the ABJ-anomaly \eqref{eq: anomaly}. Consequently, the Hopf symmetry in $\mathbb{CP}^1$ must also be non-invertible because of the symmetry matching. This non-invertibility was precisely identified by Chen and Tanizaki \cite{Chen:2022cyw,Chen:2023czk} through the non-trivial (and non-linear) structure within the Postnikov tower of $\mathbb{CP}^1$. 

C\'ordova-Dumitrescu anticipated this matching, even without the concept of non-invertible symmetry. They observed that the current $j_X$ could be improved by the (gauge-non-invariant) Chern-Simons density $j_H$. This enhancement precisely flows to the Hopf-invariant density, effectively canceling the right-hand-side of \eqref{eq: anomaly}. The TQFT part $\mathcal{A}^{q,p}$ of the topological operator \eqref{eq: Dcal} represents the gauge-invariant version of this improvement. 

For a general $N_c$, we have $N_c-1$ copies of fermions $(\rho_{i,k},\psi_{\pm,k})$, inducing the effective action \eqref{eq: effective charge} multiplied by $N_c-1$. Therefore, a Hopf soliton must have $\U(1)_X$ charge $N_c-1$, and a $\mathbb{Z}_{N_c-1}$ subgroup of $\U(1)_X$ must decouple from the IR physics. 
Consequently, the Hopf soliton is a fermion when $N_c$ is even, and a boson when $N_c$ is odd, meaning that the theta angle is $\theta = (N_c-1)\pi$. This is in addition consistent with the $\SU(2)$ global Witten anomaly \cite{Witten:1983tx,Witten:1983tw} and the recent analysis by Brennan and Intriligator \cite{Brennan:2023vsa}.
The IR Hopf soliton operator can be generated by a QED operator like 
\begin{equation}
    \prod_{k=1}^{N_c-1} \rho_{i,k} \label{eq: rhoprod}
\end{equation}
where $k$ is the index for the Cartan of $\SU(N_c)$.%

Lastly, we comment on the last term in \eqref{eq:potential}, which explicitly breaks $\U(1)_X$ and is perturbatively relevant.
When $M_\text{soft}\ll e^2\Lambda$, this term dominates and the theory is in the trivial phase, where $\U(1)_X$ is explicitly broken.
However, as \cite{Cordova:2018acb} found, the classical analysis indicates that above a critical value $M_*$ of $M_\text{soft}$, the theory flows to the $\mathbb{CP}^1$ model, where $\U(1)_X$ is restored.
This is a breakdown of the naive dimension counting diagnosis, and because of the stability of $\mathbb{CP}^1$ phase by $\SU(2)$-preserving deformations is robust against RG flows. In particular, the RG flow to $\mathbb{CP}^1$ is continuously connected to the $\U(1)_X$ preserving flow---which is depicted by the orange hollow arrow in Fig.~\ref{fig:RG_cartoon}---by turning off the last term in \eqref{eq:potential}, and therefore the matching of $\U(1)_X$ in the context of the C\'ordova-Dumitrescu flow is still meaningful despite the offending term, an instance of a \textit{dynamical symmetry restoration}.

\medskip

\noindent \textbf{Candidate for UV Baryon Operator.} In analogy to the relation between the baryons and the Skyrmions in QCD, we expect the IR Hopf soliton should be created by a ``baryon" operator in the UV. Assuming that the IR of $\text{AdQCD}_{4}^{(2)}$ is $(\mathbb{CP}^1)^{\sqcup N_c}$, what is this ``baryon" operator?
This question was asked in \cite{Bolognesi:2007ut} for $\text{AdQCD}_{4}^{(2)}$, and earlier in \cite{Bolognesi:2006ws} for QCD with fermions in (anti)symmetric representations. Bolognesi and Shifman noted that the IR soliton energy scales as $N_c^2$ in the large $N_c$ limit, and therefore a guess like (suppressing $\SU(2)$ index)
\begin{equation}
\epsilon_{j_1,j_2,\cdots,j_{N_c}}\epsilon_{t_1,t_2,\cdots,t_{N_c}} \lambda^{j_1, t_1}\cdots \lambda^{j_{N_c}, t_{N_c}},
\label{eq:bibaryon}
\end{equation}
which consists of $N_c$ quarks, is not appropriate.

In our context we can also see another problem of \eqref{eq:bibaryon}. Assuming the scenario (1a) in Fig.~\ref{fig:RG_cartoon}, after dynamical abelianization, \eqref{eq:bibaryon} reduces to an operator with non-definite $\U(1)_X$ charge, unlike \eqref{eq: rhoprod}. Here we propose a candidate baryon operator (again suppressing $\SU(2)$ indices):
\begin{equation}
    \mathcal{B} \stackrel{?}{\sim} \epsilon_{a_1,\cdots, a_{N_c^2-1}} \lambda^{a_1}\cdots \lambda^{a_{N_c^2-1}}
    = \prod_a \lambda^a
    \label{eq: Baryon}
\end{equation}
where $a_\ell = 1, \cdots N_c^2-1$ (for each $\ell$) is an index running on the adjoint representation of $\SU(N_c)$.  The gauge invariance of this operator is evident as it is indeed invariant under $\SO(N_c^2-1)$.
We can rewrite $\mathcal{B}$ as the product $\prod_k \lambda^k \prod_{\alpha\in \Delta_+} \lambda^{\alpha}\lambda^{-\alpha}$, where $\alpha$ runs through the positive roots $\Delta_+$.
Assuming that we have the condensate induced mass term $\sum_k \alpha_k \langle\phi_k\rangle \lambda^\alpha \lambda^{-\alpha}$, the products over roots in $\mathcal{B}$ is screened in the abelianized phase, resulting in \eqref{eq: rhoprod}. Moreover, $\mathcal B$ contains $\mathcal{O}(N_c^2)$ quarks, and has the correct fermion parity. 

Another way of justify the composite \eqref{eq: Baryon} %
is to embed the $\text{AdQCD}_{4}^{(2)}$ into the $\Spin(N_c^2-1)$ QCD with $n_f=2$ flavors in the vector representation.
 The pattern of continuous symmetry breaking in this theory is the same as the one in the $\text{AdQCD}_{4}^{(2)}$ and thus %
 the IR is supposed to be disjoint copies of $\mathbb{CP}^1$'s \cite{Auzzi:2008hu}.  Introducing a Higgs boson and a potential breaking the gauge symmetry to $\SU(N_c)$, obtaining $\text{AdQCD}_{4}^{(2)}$, it is natural to expect that the IR phase stays the same up to emergence of disjoint vacua.
 The baryon operator in the $\Spin(N_c^2 - 1)$ QCD is the same as \eqref{eq: Baryon}, which should flow to the Hopf soliton in the IR. 
 Thus, it is natural to expect that \eqref{eq: Baryon} also generates the same soliton after the UV theory is Higgsed to $\SU(N_c)$.

\medskip

\noindent \textbf{Conclusion and Prospects.} In this letter we have discussed the matching of an emergent non-invertible symmetry in the context of RG flows determining phases for $\text{AdQCD}_{4}^{(2)}$. Assuming $\text{AdQCD}_{4}^{(2)}$ has a dynamical abelianization in the same universality class as that of $\mathcal{N}=2$ SYM \cite{Cordova:2019jnf}, we pointed out that an emergent non-invertible $\U(1)_X$ symmetry in the abelianized description matches with IR $\mathbb{CP}^1$ phases. We also proposed a candidate of the UV baryon operator, based on the matching of the symmetry, which is expected to flow to the Hopf soliton in the IR. 

As an immediate future prospect, further exploration of the consequences of the non-invertible symmetry in $\text{AdQCD}_{4}^{(2)}$ would be of significant interest.
Specifically, the non-invertible symmetry in the IR phase leads to vortex-catalyzed decay of the Hopf soliton \cite{Chen:2022cyw}---analogous to the Callan-Rubakov monopole catalysis of proton decay (see \cite{Callan:1983sy,Rubakov:1988aq} for reviews)---that would be interesting to give a microscopic description with our methods.

We expect that emergent non-invertible symmetry matching can be exploited broadly to formulate dictionaries between UV and IR phases robust against RG flows (including dynamical symmetry restorations).  Future work could extend this analysis to other gauge theories. Whenever the assumed IR phase of a given theory has a non-invertible symmetry, suggestive of an underlying ABJ-anomalous symmetry, this  hints at a dual fermionic content, without necessarily relying on a deformed supersymmetric theory. 
Such progress would significantly advance our understanding of confinement and chiral symmetry breaking in a wider class of gauge theories, revealing how monopole condensation drives the dynamics within explicit abelian models. 

\medskip

\paragraph{Acknowledgements.---}
We thank Thomas Dumitrescu and Aleksey Cherman for comments.
The work of MDZ and DM is supported by the European Research Council (ERC) under the European Union’s Horizon 2020 research and innovation program (grant agreement No. 851931) and by the VR project grant No. 2023-05590. MDZ also acknowledges the VR Centre for Geometry and Physics (VR grant No. 2022-06593). KO is supported by JSPS KAKENHI Grant-in-Aid No.22K13969 and No.24K00522. MDZ, SNM, DM and KO also acknowledges support from the Simons Foundation Grant \#888984 (Simons Collaboration on Global Categorical Symmetries).

\bibliography{references}

\clearpage

\appendix

\section{End Matter}

\medskip

\noindent \textbf{On the breaking of the accidental $SU(2)^{N_c-1}$ symmetry.} In the main text, in the context of the deformed $\mathcal{N}=2$ $
\SU(N_c)$ SYM theory, we assumed a flow from the 
multimonopole point EFT to the $\mathbb{CP}^1$ sigma model.
In this End Matter we elaborate on how this flow should occur, when $N_c>2$. Our result is consistent with the D'Hoker, Dumitrescu, Gerchkovitz, and Nardoni cascade \cite{Nardoni:talk,DumitrescuNardoni:talk,DHoker:unpublished}.

The multimonopole point is descried by a tensor product of copies of SQED, which we denote SQED${}^{\otimes(N_c-1)}$.
Via a naive generalization of the C\'ordova-Dumitrescu analysis, 
one might expect to flow to the tensor product of $\mathbb{CP}^1$'s, $(\mathbb{CP}^1)^{\otimes (N_c-1)}$.\footnote{
The tensor product should not be confused with the disjoint union 
of $\mathbb{CP}^1$'s which refers to disconnected vacua, coming 
from discrete symmetry breaking.}
However, by naturalness, we should consider all the deformations 
of the tensor product theory that preserves the symmetry of the UV 
theory. Among the $\SU(2)^{N_c-1}$ symmetry acting on $(\mathbb{CP}^1)^{\otimes (N_c-1)}$, the only $SU(2)$ symmetry 
existing in the UV non-Abelian gauge theory is its diagonal, identified with $\SU(2)_R$.  

Among the deformations preserving the diagonal $\SU(2)$, there 
are relevant ones; if we denote $\Phi_k(x)$ the field valued in the $k$-the copy of $\mathbb{CP}^1$ when evalued at the spacetime point $x$, then one of  the relevant terms is the distance, measured on $\mathbb{CP}^1$, between $\Phi_{k_1}(x)$ and $\Phi_{k_2}(x)$ with some indices $k_1$ and  $k_2$.
If the sign of the potential is so that the points $\Phi_{k}(x)$ wants to coincide, the term triggers a flow from $(\mathbb{CP}^1)^{\otimes (N_c-1)}$ to $\mathbb{CP}^1$.

However, it is not straightforward to understand the corresponding term breaking $\SU(2)^{N_c-1}$ in the SQED${}^{\otimes(N_c-1)}$ description.
The supersymmetry breaking deformation in the UV $\SU(N_c)$ is the bottom 
component of an (improved) stress-energy tensor, and hence we expect the 
same for the corresponding deformation at the multimonopole point. One
 such term does not contain an $\SU(2)^{N_c-1}$ breaking term, however, and 
 so at the first order $M_\text{soft}$ such symmetry breaking effect is absent. 
 We have to look at more subtle effect. The SQED${}^{\otimes(N_c-1)}$ is the 
 EFT of the multimonopole point, and thus contains marginal or irrelevant  deformations other than the bottom component of stress-energy tensor.
These are induced upon integrating out the massive particles, e.g.\ the W-bosons, and in particular either preserve or break the supersymmetry.\footnote{The analysis of \cite{Nardoni:talk,DumitrescuNardoni:talk,DHoker:unpublished} indeed explicitly finds such quartic potential terms, both supersymmetry breaking and preserving ones, are induced.} The latter terms are suppressed by factors of $M_\text{soft}$. We do not expect that all such terms preserve $\SU(2)^{N_c-1}$, because the symmetry is absent in the UV.
Then, such a marginal or irrelevant term naturally flows to the \emph{relevant} terms in $(\mathbb{CP}^1)^{\otimes (N_c-1)}$. This phenomenon is called dangerous deformation \cite{Fisher}, resulting in an RG flow that almost reach the $(\mathbb{CP}^1)^{\otimes (N_c-1)}$ point but eventually diverges from it turning to $\mathbb{CP}^1$. In short, $\mathbb{CP}^1$ is robust among $\SU(2)_R$ preserving phases, while $(\mathbb{CP}^1)^{\otimes (N_c-1)}$ is not.

\medskip

\noindent \textbf{$SU(2)_R$ quantum numbers of the Hopf soliton.} We conclude the End Matter with a remark about the $SU(2)_R$ charge of the Hopf soliton. The operator $\prod_{k=1}^{N_c -1} \rho_{i,k}$ of \eqref{eq: rhoprod} transforms in the $\mathbf{2}\otimes ... \otimes \mathbf{2}$ representation of $\SU(2)^{N_c-1}$. Upon breaking the symmetry to $SU(2)_{\text{diag}}$, we expect the smallest representation occuring gives $SU(2)_\text{diag}$ quantum numbers to the ground state of the Hopf soliton. The latter therefore transforms in a doublet of  $SU(2)_\text{diag}$ when $N_c$ is even, and in a singlet when $N_c$ is odd.

\end{document}